\documentclass[a4paper,11pt]{article}
\pdfoutput=1 % if your are submitting a pdflatex (i.e. if you have
\usepackage{epsfig}
\usepackage{graphicx}                  % provides \includegraphics command
\usepackage{ae}
\usepackage{amsmath}
\usepackage{amssymb}
\usepackage{graphics}
\usepackage{dcolumn}
\usepackage{bm}
\usepackage{lineno}
\usepackage{caption}
\usepackage{ragged2e}
\renewcommand{\thefootnote}{\fnsymbol{footnote}}
\title{\bf Investigation of the stability in the performance of triple GEM detectors for High Energy Physics experiments}
\date{}

\begin{document}

\maketitle
	\flushbottom
\vspace*{-1cm}
\centering

\author{\bf S.~Mandal$^{1}$,}
\author{\bf S.~Chatterjee$^{1^a}$,}
\let\thefootnote\relax\footnotetext{$^a$Now at University of Massachusetts, Amherst, USA.} 
\author{\bf A.~Sen$^{1^b}$,}
\let\thefootnote\relax\footnotetext{$^b$Now at Ohio University, Athens, USA.}
\author{\bf S.~Gope$^{1}$,}
\author{\bf S.~Dhani$^{2}$,}
\author{\bf A.~C.~Hegde$^{3}$,}
\author{\bf M.~Chatterjee$^{4}$,}
\author{\bf S.~Das$^{1}$,}
\author{\bf S. Biswas$^{1^*}$}
\let\thefootnote\relax\footnotetext{$^*$Corresponding author. 

\hspace*{0.4cm}E-mail: saikat@jcbose.ac.in }

\vspace*{0.5cm}

	$^1${{Department of Physical Sciences, Bose Institute, EN-80, Sector V, Kolkata-700091, India}
	
	$^2${{Department of Physics, IIT Bombay, Powai, Mumbai, Maharashtra 400076, India}
	
	$^3${ {School Of Physical Sciences, NISER, Jatni, Odisha 752050, India}
	
	$^4${ {Department of Physics, St. Xavier's College (Autonomous), 30, Mother Teresa Sarani, Kolkata - 700016, West Bengal, India}

\vspace*{0.5cm}
\centering{\bf Abstract}
\justify

Gas Electron Multiplier (GEM) is one of the mostly used gaseous detectors in the High Energy Physics (HEP) experiments. GEMs are widely used as tracking devices due to their high-rate handling capability and good position resolution.

An initiative is taken to study the stability in performance of the GEM chamber prototypes in the laboratory using external radiation for different Argon based gas mixtures. The effect of ambient parameters on the gain and energy resolution are studied. Very recently some behavioural changes in the performance of a SM GEM chamber is observed. The details of the experimental setup, methodology and results are reported here.



%\begin{keyword}
%% keywords here, in the form: keyword \sep keyword, up to a maximum of 6 keywords
%keyword 1 \sep keyword 2 \sep keyword 3 \sep keyword 4



%\end{keyword}
\vspace*{0.25cm}


%\tableofcontents



%\linenumbers
\section{Introduction} \label{intro}

The advancement of accelerator technologies helps the High Energy Physics (HEP) community to reach higher collision rates to measure rare physical observables with unprecedented precessions \cite{Galatyuk}. This imposes a great challenge to the rate handling capabilities of the detectors. In the HEP experiments, gas-filled detectors are generally used for tracking, triggering, and timing measurements. Gas Electron Multiplier (GEM) is the most commonly used detector among the Micro-Pattern Gas Detector (MPGD) family. This detector, introduced by Prof. F. Sauli in the year 1997 \cite{Sauli}, has become popular in HEP experiments because of its good position resolution ($\sim$~100~$\mu m$) and very high rate handling capability ($\sim$~1 MHz/mm$^2$) \cite{Sauli, Buzulutskov, Ketzer}. Future Heavy-Ion (HI) experiment CBM (Compressed Baryonic Matter) at FAIR, Darmstadt, Germany also plans to use triple GEM detector for the Muon Chambers (MuCh) which is expected to be operated under high particle flux \cite{cbm, fair, rama}. 

Long term stability in performance is one of the main criteria for choosing detectors for the HEP experiments. Investigating the performance of the chambers with prolonged irradiation is one of the useful methods to understand the stability of the chambers. Thus, an initiative is taken to study the performance of different gas-filled detectors in the laboratory using external irradiation for different gas mixtures. This R\&D activity includes stability and rate-handling capability study of straw-tube detector, development of Resistive Plate Chamber (RPC) using indigenous bakelite plate applying a new technique of linseed oil coating and study of performance of GEM detector \cite{Roy_thesis, Chatterjee_thesis, Sen_thesis}. This R\&D of GEM detector includes study of uniformity in performance over the active area, stability test of gain and energy resolution, study of radiation induced effects such as charging-up on the performance for both double mask (DM) and single mask (SM) GEM chambers \cite{adak, s_roy_gain_calculation, uniformity_1, chatterjee_2020, s_chatterjee_charging_up_1, s_chatterjee_charging_up_2, chatterjee_2023_rh, chatterjee_2023_charge}.

%A typical GEM foil is made up of copper cladding (5 $\mu m$) and Kapton foil (50 m). One of the most useful techniques is the photolithographic process, which is used to etch a large number of holes into the foil. The foil is generally divided into two types: depending on the etching technique, one is a single mask, and the other is a double mask. For the manufacturing of large-area GEM chambers, the single-mask approach is used over the double-mask technique. In contrast to the holes made using the double mask technique, the holes made using the single mask technique have an uneven bi-conical shape. Numerous experiments are carried out to understand how hole shape modification affects the final performance of the chamber. The dielectric medium (Kapton) in the detector's active volume changes the behaviour of the chamber when it is subjected to external radiation. Consequently, the gain of the chamber increases initially and eventually approaches a constant value. 

For the stability test of GEM detectors a $^{55}$Fe X-ray source is used to irradiate the chambers. The same source is also used to monitor the spectrum. The effect of temperature, pressure, relative humidity on the gain and energy resolution and the charging up effects are studied. Very recently some behavioural changes in the performance of a SM triple GEM chamber is observed and the same is being reported in this article.

%In the present study, an investigation has been made to study the performance of the GEM chamber prototypes in the laboratory using external irradiation for different gas mixtures.

%In the present study, an investigation has been made to study the performance of the GEM chamber prototypes in the laboratory using external irradiation for different gas mixtures.

%Here is where you provide an introduction to work and some background. For example building on previous work of image enhancment in optical astronomy \citep{vojtekova2021learning}, \cite{sweere2022deep} developed a method to improve the resolution of X-ray images from XMM-Newton to provide similar spatial resolution to Chandra.

\section{Detector configuration and experimental setup} \label{det_con}
%%\label{}
%%\lipsum[1]

The detector prototype used in this work is a SM triple GEM in 3-2-2-2 configuration (drift gap - transfer gap 1- transfer gap 2 - induction gap), having a dimension of 10~$\times$~10~cm$^2$. The GEM layers and drift plane are biased using a resistive chain network as shown in Figure~\ref{fig1}. In this network 10~M$\Omega$ protection resistors are used in the drift plane and top layers of each GEM foil. 1~M$\Omega$ resistors are connected in the drift, transfer and induction gaps whereas 560~k$\Omega$ are used across each GEM foil as shown in Figure~\ref{fig1}. Although there are 256 X and 256 Y segments on the readout pad, but in this whole study, the signal is taken using sum-up boards instead of the individual readout from the XY strips. One sum-up can accommodate either 128~X or 128~Y strips. Four such sum-up boards are used for this chamber.

%%%%%%%%%%%%%%%%%%%%%%%%%%%%%%%%%%%%%%%%%%%%%%%%%%%
	\begin{figure}[htb!]
		\begin{center}
			\includegraphics[scale=0.25]{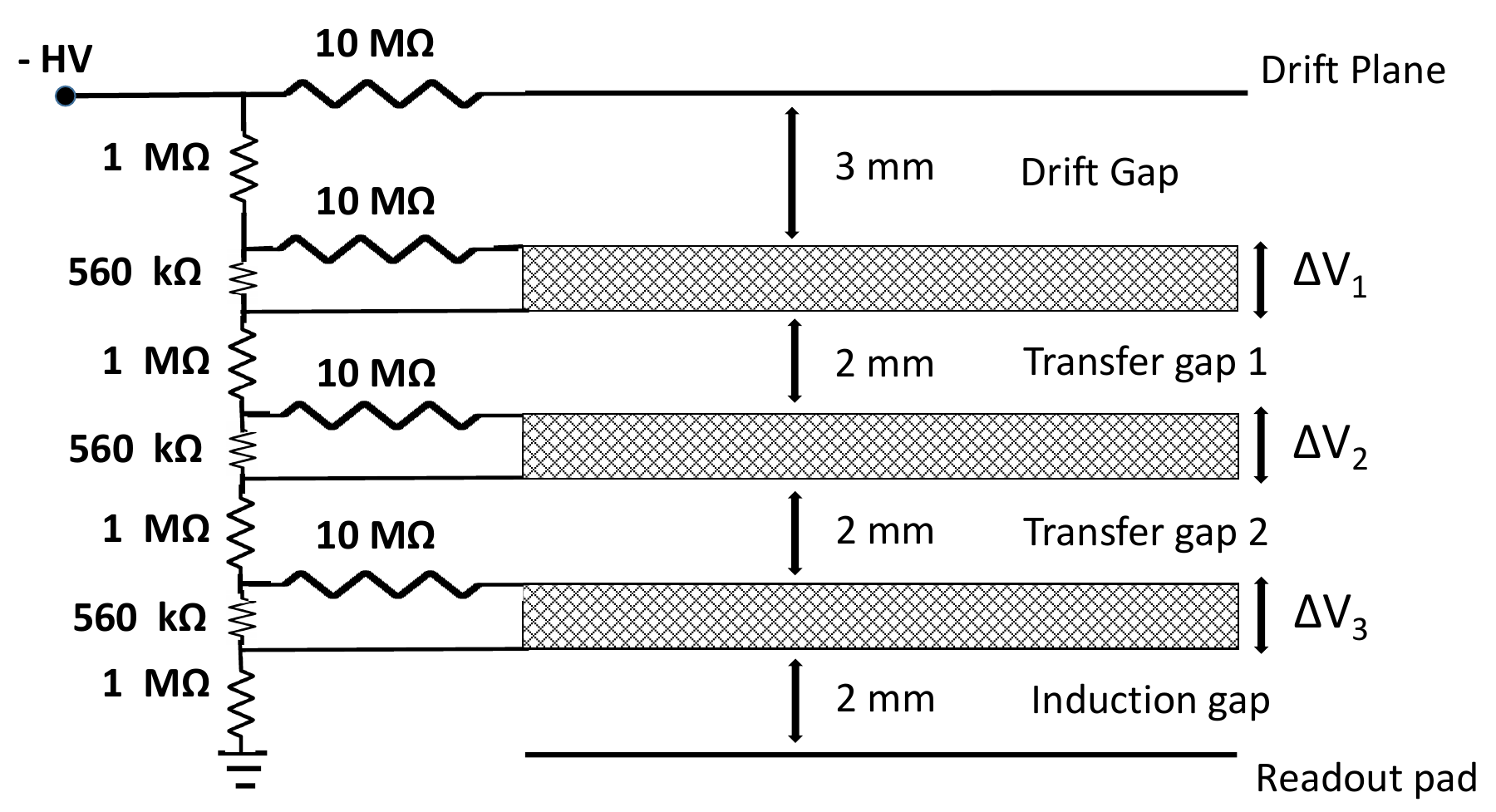}
			\caption{\label{fig1} Schematic of the high voltage (HV) distribution through the resistive chain network to different planes of the SM triple GEM detector~\cite{s_chatterjee_charging_up_2}. }\label{fig1}
		\end{center}
	\end{figure}
%%%%%%%%%%%%%%%%%%%%%%%%%%%%%%%%%%%%%%%%%%%%%%%%%%%

For the result presented here, a mixture of Argon (Ar) and CO$_2$ in 70/30 volume ratio is used. A constant 2~l/h flow rate is maintained using a V{\"o}gtlin gas flow meter. In order to expose a specific circular patch of the detector to the X-ray with a characteristic energy of 5.9~keV from the $^{55}$Fe source, a circular collimator having diameter 8~mm is used for all measurements, which corresponds to an area of exposure of about 50~mm$^2$ on the detector.

The signal from the sum-up board is fed to a charge sensitive preamplifier (VV50-2) having a gain of 2~mV/fC and shaping time of 300~ns~\cite{preamplifier}. The output from the preamplifier is placed in a linear Fan-In Fan-Out (FIFO) module. One output from the FIFO is fed to a Multi Channel Analyzer (MCA) to store the X-ray spectra in the desktop computer. Another output from the FIFO is fed to a Single Channel Analyzer (SCA), the output of which above the noise threshold is counted using a NIM scaler. The threshold to the SCA is kept constant at 0.4~V throughout the measurements. A more detail description of the experimental setup is available in Ref~\cite{s_chatterjee_charging_up_2}.

\section{Mathematical formalism} \label{math}

As mentioned in section~\ref{det_con}, to irradiate the chamber, a $^{55}$Fe X-ray source is employed. Figure~\ref{fig22} displays a typical $^{55}$Fe X-ray spectrum at an high voltage (HV) of -~4600~V, which corresponds to a $\Delta{V}$ of $\sim$~396.2~V across each of the GEM foils. The 5.9~keV main peak and 2.9~keV Argon escape peak along with the noise peak are clearly visible in the spectrum. The gain and energy resolution of the chamber is calculated by fitting the  5.9~keV main peak with a Gaussian distribution. The mean of the fitted Gaussian spectra gives the information of the total output charge by using  the gain of the preamplifier and the MCA calibration factor.

%%%%%%%%%%%%%%%%%%%%%%%%%%%%%%%%%%%%%%%%%%%%%%%%%%%
	\begin{figure}[htb!]
		\begin{center}
	\includegraphics[scale=0.40]{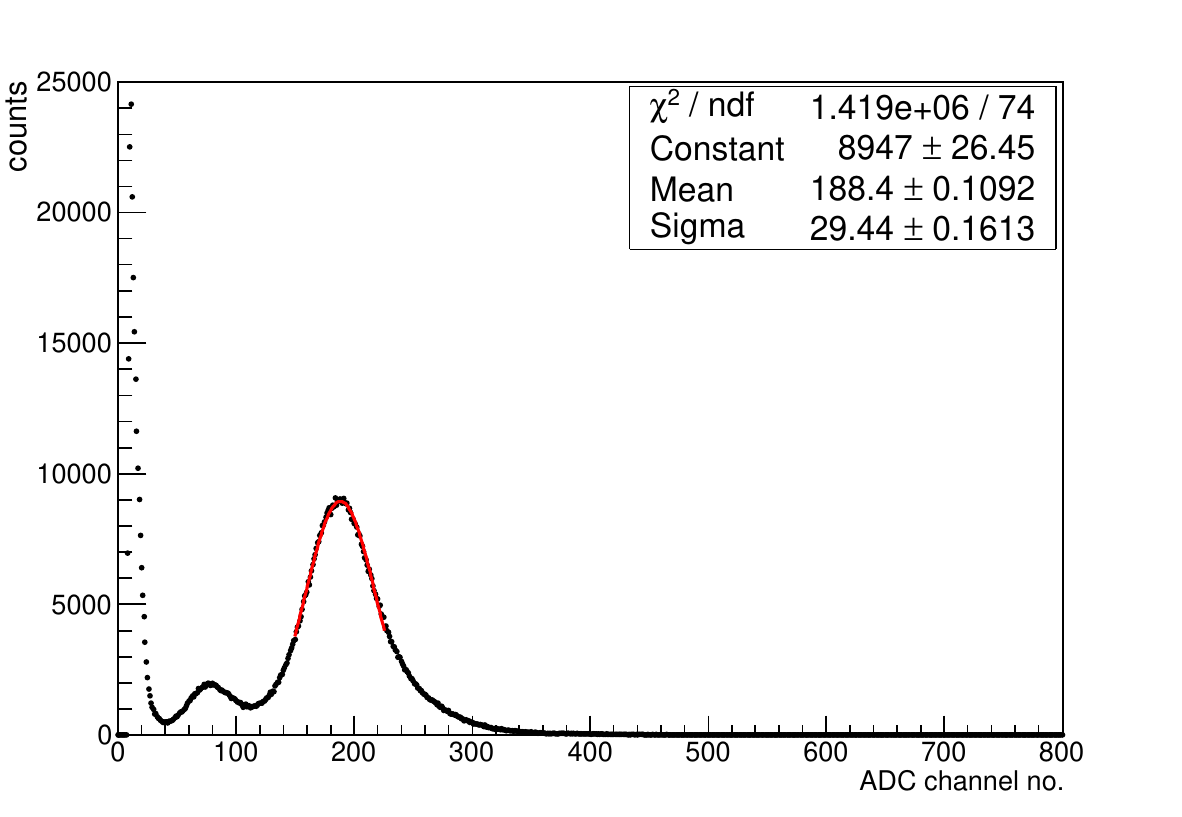}
	\caption{(Colour online) Typical $^{55}$Fe spectrum at - 4600~V. The $\Delta V$ across each of the GEM foils is $\sim$~396.2~V.}\label{fig22}
\end{center}
	\end{figure}
%%%%%%%%%%%%%%%%%%%%%%%%%%%%%%%%%%%%%%%%%%%%%%%%%%%

%\begin{figure}
%\centering 
%\includegraphics[width=0.4\textwidth]{electronic.png}		
%\caption{Schematic of the electronic circuit used for data acquisition.}
%\label{fig1}%
%\end{figure}

%%\subsection{Subsection title}

%\begin{figure}
%	\centering 
%	\includegraphics[width=0.4\textwidth, angle=-90]{Results_in_Physics.pdf}	
%%	\label{fig_mom0}%
%\end{figure}

%A random equation, the Toomre stability criterion:

%\begin{equation}
%    Q = \frac{\sigma_v \times \kappa}{\pi \times G \times \Sigma}
%\end{equation}

According to the definition, the mathematical formula of gain of the detector is given by

\begin{equation}
\begin{aligned}
gain  
& = \frac{output \; charge}{input \: charge} & = \frac{\frac{Pulse \; hight}{2 \: mV} fC}{no.\; of \; primary \; electrons\; \times \; eC} 
\end{aligned}
\end{equation}

The Pulse height in Volt is found out using the MCA calibration factor. The MCA is calibrated using a known pulse height from a pulse generator and the relation between the pulse height and the mean MCA channel number can be expressed as,

\begin{equation}
\begin{aligned}
Pulse \; height \; (V) & = MCA \; Channel \; no. \times 0.0014 \; + \;  0.1428
\end{aligned}
\end{equation}
The details of the MCA calibration is described in Ref~\cite{Chatterjee_thesis}.

The full width at half maxima (FWHM) of the Gaussian fitted spectra is used to define the energy resolution of the chamber. Energy resolution is calculated using the relation,

\begin{equation}
\centering
Energy \; resolution = \frac{sigma \times 2.355}{mean}   
\end{equation}
where the $sigma$ and the $mean$ are obtained from the Gaussian fitting of the energy spectra.

For stability test of any GEM detector the spectra and the ambient parameters such as temperature ($t$ in $^\circ$C), pressure ($p$ in mbar) and relative humidity ($RH$ in \%) are recorded with finite time interval. The gain ($G$) of gaseous detector depends significantly on the ratio of absolute temperature ($T$ = $t$+273) and pressure ($p$), according to the relation,

\begin{equation}
\centering
G (T/p) = A e^{(B \frac{T}{p})}\label{correlation}
\end{equation} 
Where, $A$ and $B$ are constants.

To check the stability in gain, the gain vs. $T/p$ correlation plot is fitted with the function given by Eq.~\ref{correlation} and the values of the fit parameters A and B are extracted. The measured gain is then normalised with the gain calculated from Eq.~\ref{correlation} using the formula 

\begin{equation}
\centering
normalised \; gain = \frac{measured \; gain}{A e^{(B \frac{T}{p})}}\label{norm_g} 
\end{equation}

Finally, the normalised gain is plotted as a function of the total charge accumulated per unit irradiated area of the GEM chamber, which is directly proportional to time. The charge accumulated per unit area ($dq$) of the detector at a particular time is calculated by

\begin{equation}
\centering
\frac{dq}{dA} = \frac{r \times n \times e \times G \times dt}{dA}\label{chperarea}
\end{equation} 

where, $r$ is the measured rate in Hz incident on a particular area of the detector, $dt$ is the time in second, $n$ is the number of primary electrons for a single X-ray photon, $e$ is the electronic charge, $G$ is the gain and $dA$ is the area of the irradiated portion.

Similarly the correlation curve between the energy resolution and $T/p$ is also fitted with the exponential function:

 \begin{equation}
 \centering
 energy \; resolution = A' e^{(B' \frac{T}{p})}\label{correlation_er}  
 \end{equation}

Where, $A'$ and $B'$ are two constant fit parameters.

The measured energy resolution is then normalised with the energy resolution as calculated from Eq.~\ref{correlation_er} using the formula

\begin{equation}
\centering
%\begin{aligned}
normalised \; energy \; resolution = 
 \frac{measured \; energy \; resolution}{A' e^{(B' \frac{T}{p})}}\label{norm_er} 
%\end{aligned}
\end{equation}
 
\section{Results}\label{res} 

Results of stability test of the SM triple GEM detector is described here. Initially the HV is set at -~4500~V and the spectra are recorded for every 1 minute. The t, p and RH are also recorded for every 1 minute interval using a data logger made in-house \cite{sahu}. The divider current is also recorded time to time from the HV power supply itself. The applied voltage and the divider current as a function of time is shown in Figure~\ref{fig3} and the variation of measured gain, \% energy resolution and T/p are shown in Figure~\ref{fig4}. 

%%%%%%%%%%%%%%%%%%%%%%%%%%%%%%%%%%%%%%%%%%%%%%%%%%%
\begin{figure}[htb!]
	\begin{center} 
	\includegraphics[scale=0.2]{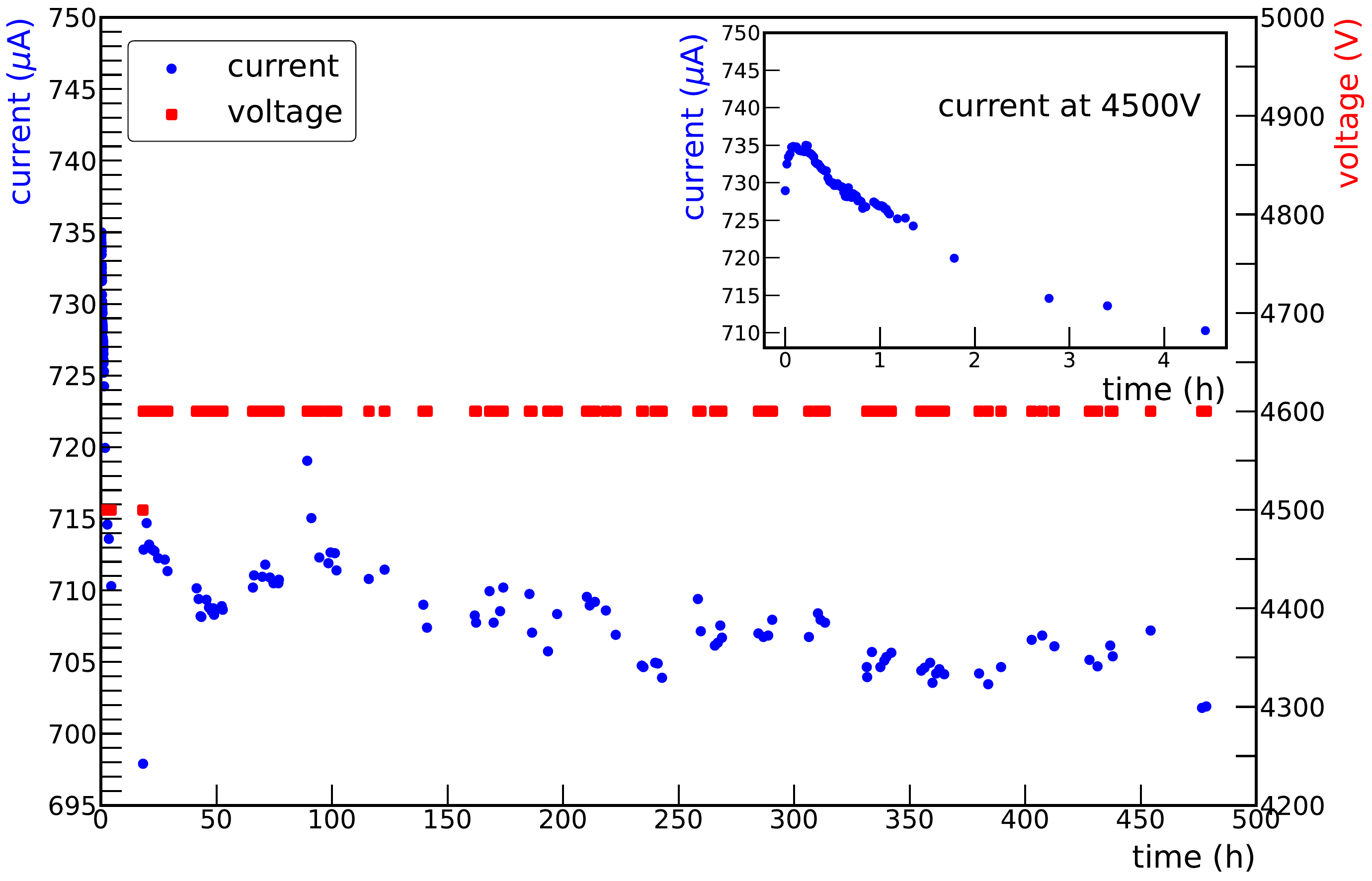}
	\caption{(Colour online) Applied voltage and divider current as a function of time. The current as a function of time for the first 5 hours after applying high voltage is shown inset.}\label{fig3}
	\end{center}
\end{figure}
%%%%%%%%%%%%%%%%%%%%%%%%%%%%%%%%%%%%%%%%%%%%%%%%%%%

%%%%%%%%%%%%%%%%%%%%%%%%%%%%%%%%%%%%%%%%%%%%%%%%%%%
\begin{figure}[htb!]
	\begin{center} 
	\includegraphics[scale=0.20]{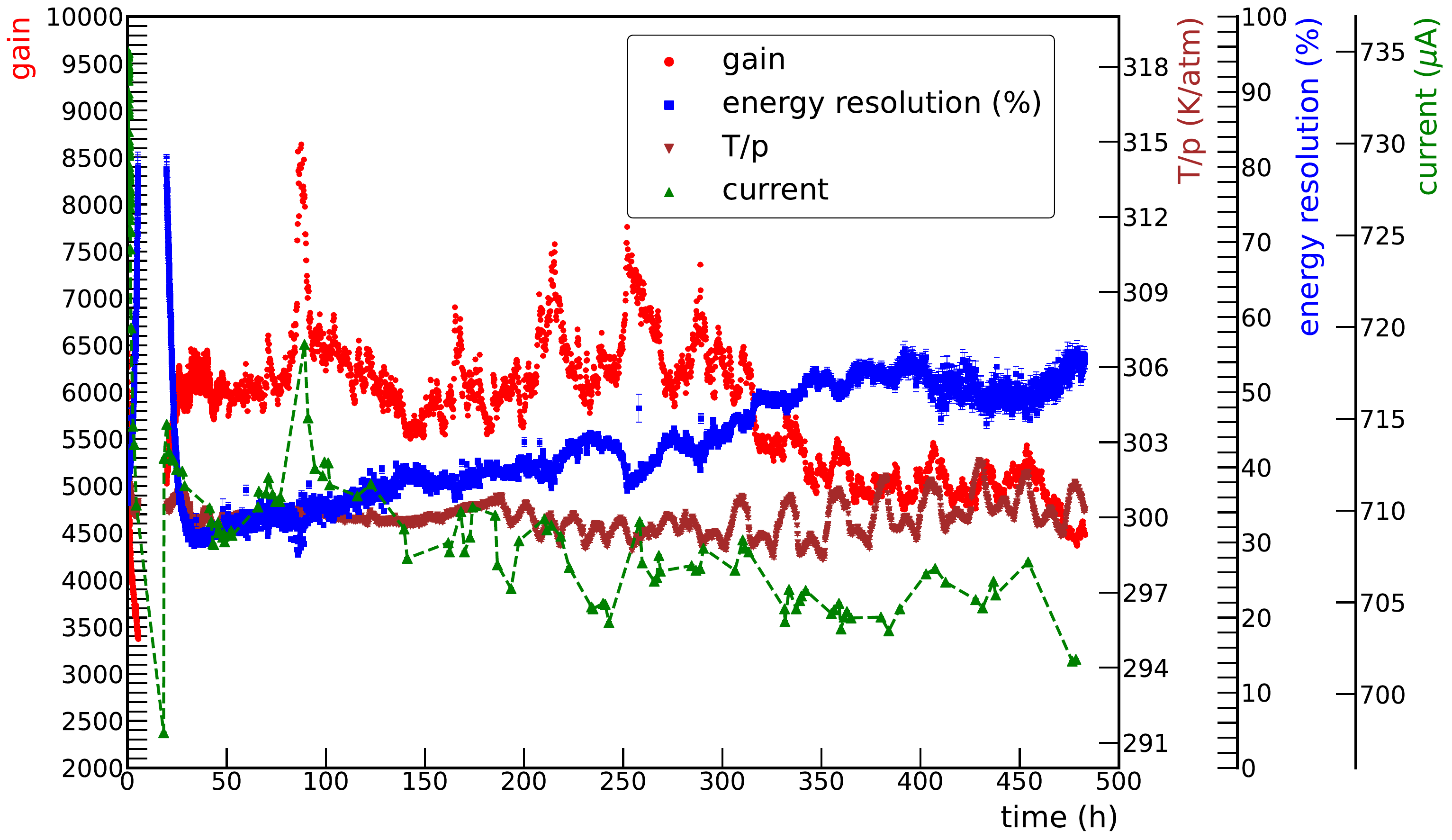}
	\caption{(Colour online) Variation of the measured gain, energy resolution and T/p as a function of the time.}\label{fig4}
	\end{center}
\end{figure}
%%%%%%%%%%%%%%%%%%%%%%%%%%%%%%%%%%%%%%%%%%%%%%%%%%%

%%%%%%%%%%%%%%%%%%%%%%%%%%%%%%%%%%%%%%%%%%%%%%%%%%%
\begin{figure}[htb!]
	\begin{center} 
	\includegraphics[scale=0.5]{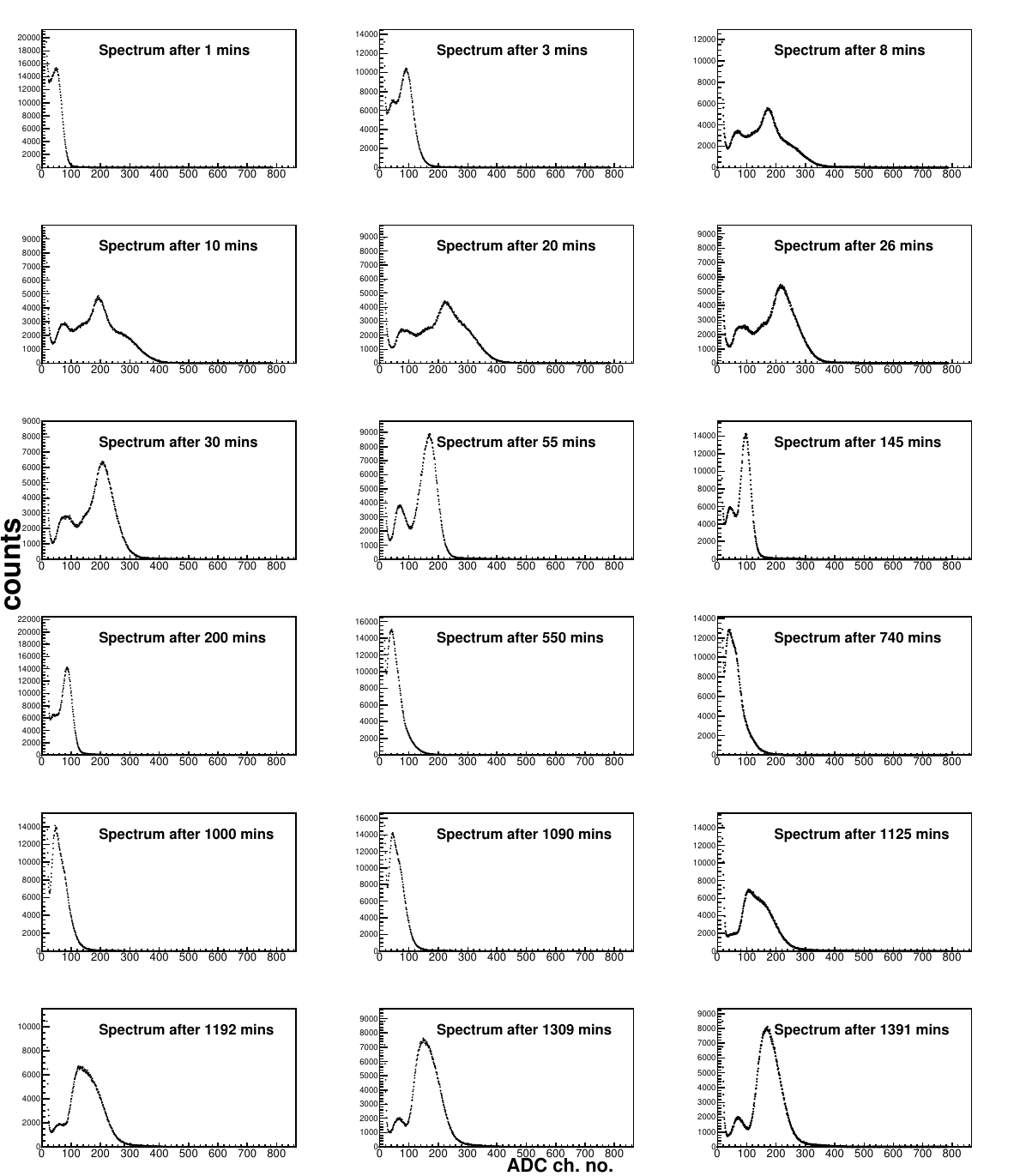}
	\caption{The shape of $^{55}$Fe X-ray spectra at different times after the application of HV. HV from -~4500~V to -~4600~V is raised after 1190 minutes from the beginning.}\label{fig9}
	\end{center}
\end{figure}
%%%%%%%%%%%%%%%%%%%%%%%%%%%%%%%%%%%%%%%%%%%%%%%%%%%

It is seen from Figure~\ref{fig3} that immediately after applying HV the current increases rapidly, reaches a maximum and then starts to decrease. The shape of $^{55}$Fe X-ray spectra recorded after different time from the application of HV are shown in Figure~\ref{fig9}. It is seen from Figure~\ref{fig9} that well shaped main and escape peaks of $^{55}$Fe spectra are obtained only after 55 minutes of the application of HV. Before that, even the well separated escape peaks are not visible at all. The main peak is fitted to calculate the gain only after 30 minutes from the start of operation, when at least a main peak is observed, but after 360 minutes it was impossible to fit the peak with the Gaussian function and as a consequence it was difficult to calculate the gain. Immediately after applying HV as the current increases rapidly and peak also shifts in the right which means the gain also increased and when the current started to decrease the gain also decreased with time as seen in Figure~\ref{fig9}. About 24 hours the current decreases which in turn decreases the gain of the chamber. The energy resolution changes on the opposite way. Again good spectra, which can be fitted by Gaussian function, are obtained after increasing the voltage from -~4500~V to -~4600~V after 1190 minutes from the beginning of first operation. The current at the starting point (at HV of -~4500~V) is measured to be about 729~$\mu$A which increases to a maximum value of 735~$\mu$A and starts to decrease. After 24 hour of operation the current decreased to about 697~$\mu$A and  the applied HV is increased to -~4600~V. At that point, the divider current is again raised to a value $\sim$~715~$\mu$A. Subsequently, as time passes, the current drops slowly. Since the current decreases and fluctuates with time, the gain also decrease and fluctuate, and the same trend is observed in the Figure~\ref{fig3} and  \ref{fig4}. On the other hand the energy resolution becomes worse as the gain decreases. After 24 hours from stating point when the HV is increased to -~4600~V the gain increases and then saturates.

The correlation plot, i.e. the measured gain is plotted as a function of T/p and fitted with the functional form shown in equation~\ref{correlation} and is illustrated in Figure~\ref{fig5}. The values of the fit parameters $A$ and $B$ are found to be 8.29~$\times$~10$^1$~$\pm$~1.64~$\times$~10$^{-1}$ and 1.36~$\times$~10$^{-2}$~$\pm$~6.34~$\times$~10$^{-6}$~atm/K respectively. Similarly another correlation plot, i.e. the measured energy resolution is plotted as a function of T/p and fitted with a function \ref{correlation_er} and is shown in Figure~\ref{fig6}. In this case, the values of $A'$ and $B'$ obtained from the fitting are  1.57~$\times$~10$^3$~$\pm$~1.74~$\times$~10$^1$ and -1.14~$\times$~10$^{-2}$~$\pm$~3.68~$\times$~10$^{-5}$ atm/K respectively.
 
%%%%%%%%%%%%%%%%%%%%%%%%%%%%%%%%%%%%%%%%%%%%%%%%%%%
\begin{figure}[htb!]
	\begin{center} 
	\includegraphics[scale=0.45]{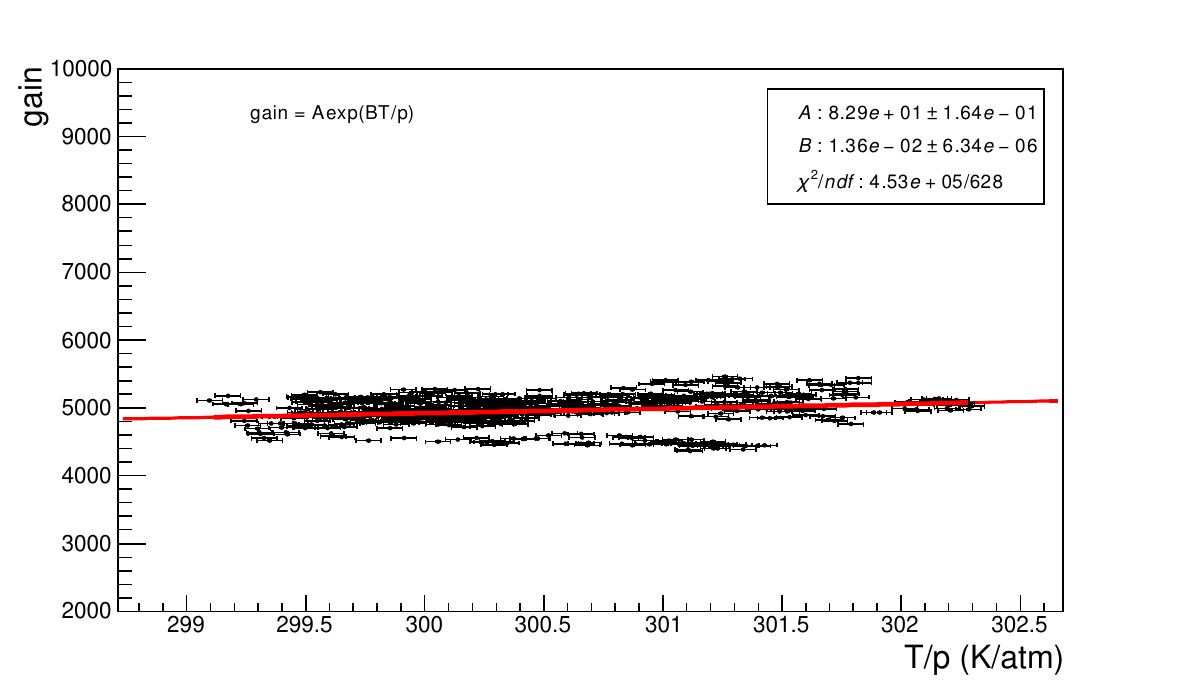}
	\caption{(Colour online) Variation of the gain as a function of T/p.}\label{fig5}
	\end{center}
\end{figure}
%%%%%%%%%%%%%%%%%%%%%%%%%%%%%%%%%%%%%%%%%%%%%%%%%%%
%\vspace{-0.6cm}

%%%%%%%%%%%%%%%%%%%%%%%%%%%%%%%%%%%%%%%%%%%%%%%%%%%
 \begin{figure}[htb!]
	\begin{center} 
 	\includegraphics[scale=0.45]{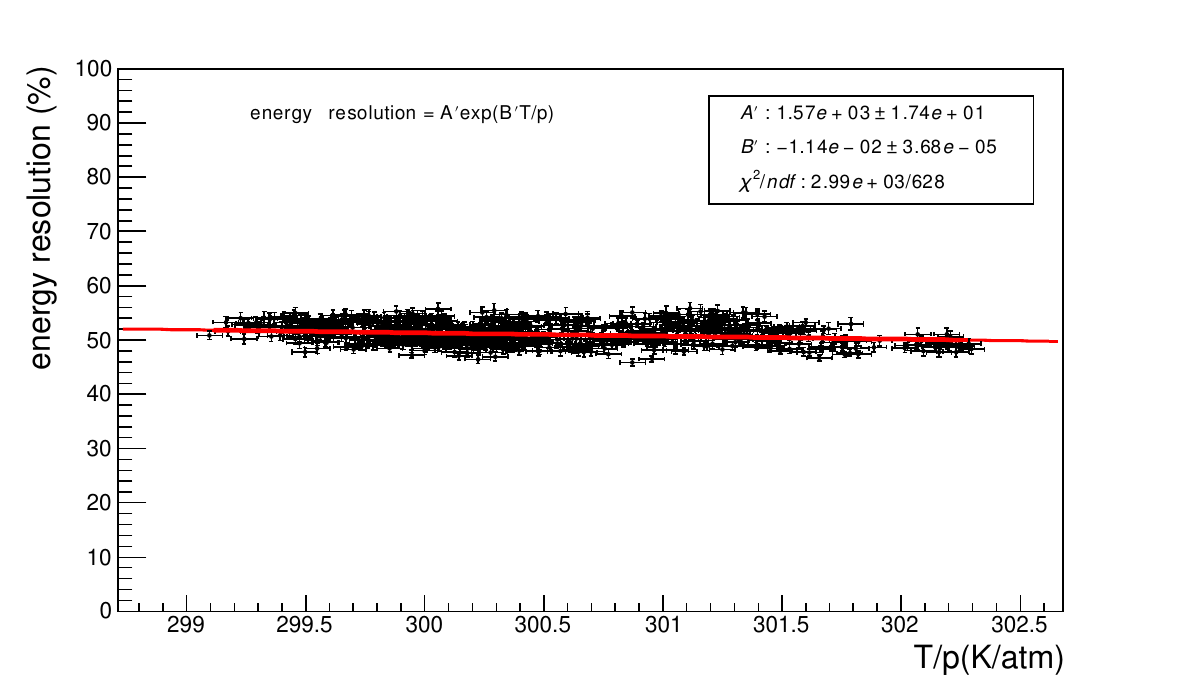}
 	\caption{(Colour online) Variation of the energy resolution as a function of T/p. }\label{fig6}
	\end{center}
 \end{figure}
 %%%%%%%%%%%%%%%%%%%%%%%%%%%%%%%%%%%%%%%%%%%%%%%%%%%

The measured gain and energy resolution are normalised using the formulae \ref{norm_g} and \ref{norm_er} respectively and the normalised gain and normalised energy resolution are plotted as a function of time in Figure~\ref{fig7}. In Figure~\ref{fig10} the normalised gain and normalised energy resolution are also plotted as a function of charge accumulated per unit area as calculated using equation~\ref{chperarea}. The average normalised gain and average normalised energy resolution are found to be 1.17~$\pm$~0.15 and 0.81~$\pm$~0.19 respectively after a charge accumulated per unit area of 1.4~mC~mm$^{-2}$.

%%%%%%%%%%%%%%%%%%%%%%%%%%%%%%%%%%%%%%%%%%%%%%%%%%%
\begin{figure}[htb!]
	\begin{center} 
	\includegraphics[scale=0.20]{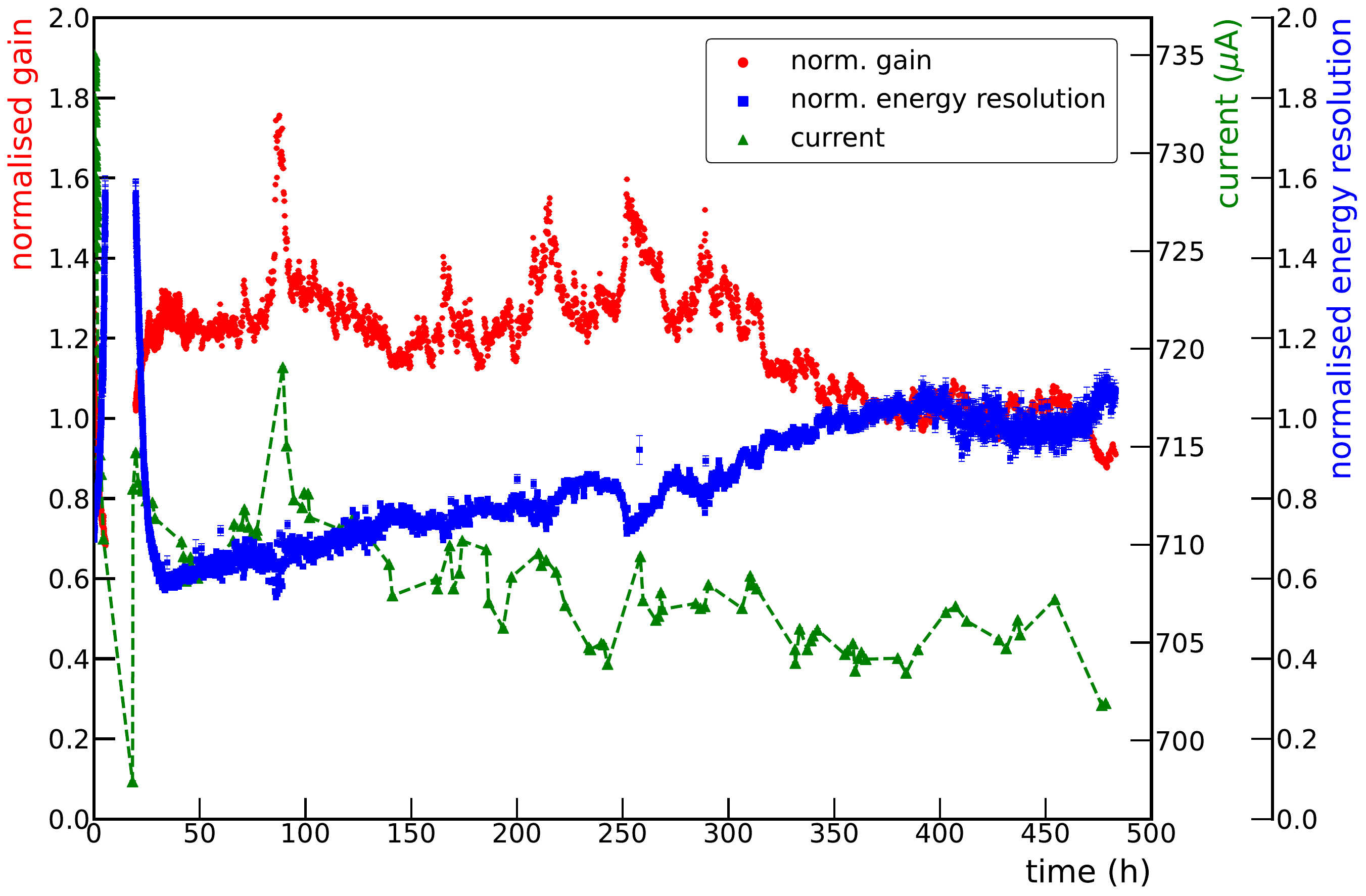}
        \caption{(Colour online) Variation of normalise gain, normalised energy resolution along with the variation of divider current as a function of time.}\label{fig7}
	\end{center}
\end{figure}
%%%%%%%%%%%%%%%%%%%%%%%%%%%%%%%%%%%%%%%%%%%%%%%%%%%
%%%%%%%%%%%%%%%%%%%%%%%%%%%%%%%%%%%%%%%%%%%%%%%%%%%
\begin{figure}[htb!]
	\begin{center} 
	\includegraphics[scale=0.20]{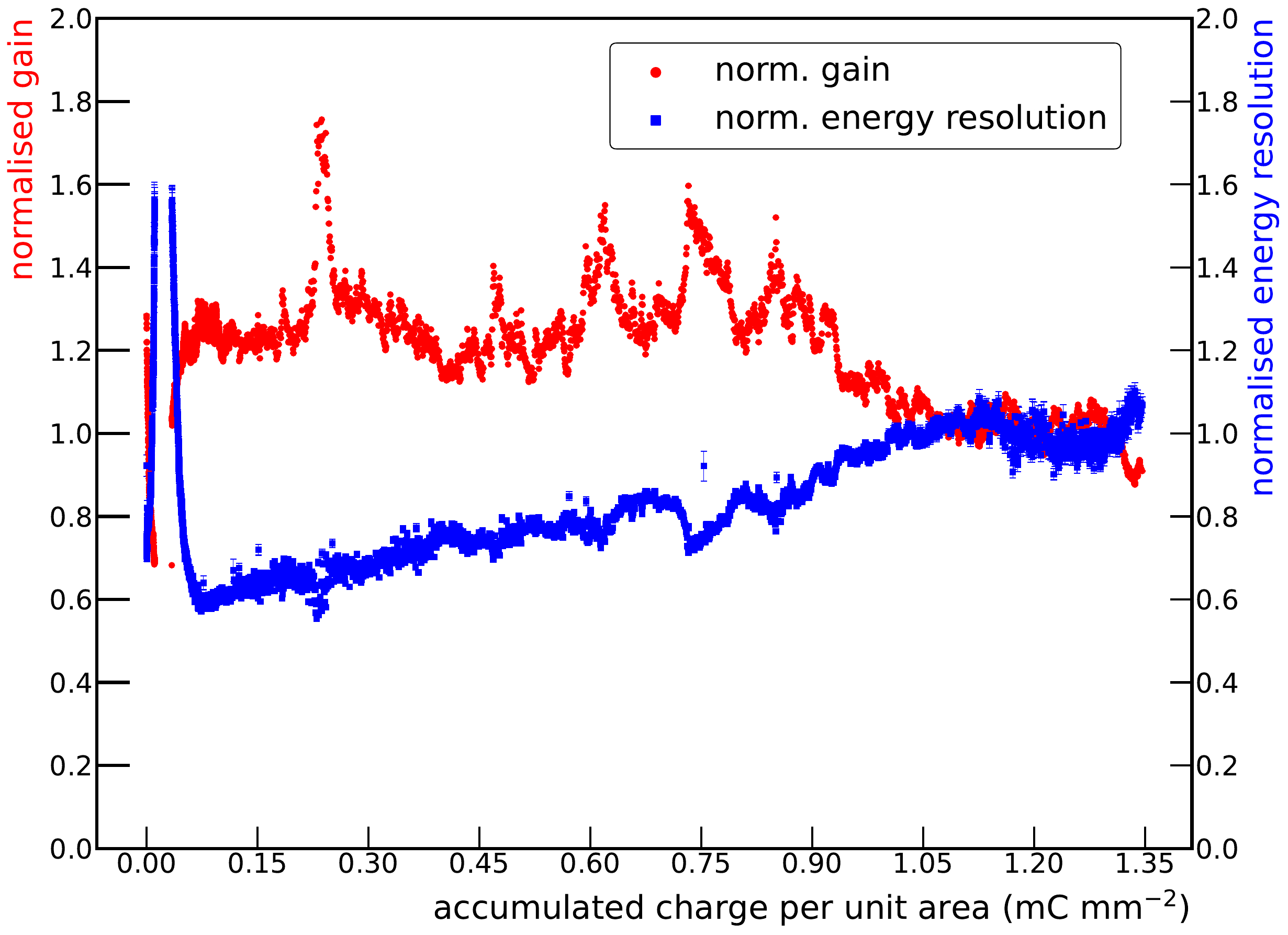}
        \caption{(Colour online) Variation of normalise gain, normalised energy resolution along with the variation of divider current as a function accumulated charge per unit area.}\label{fig10}
	\end{center}
\end{figure}
%%%%%%%%%%%%%%%%%%%%%%%%%%%%%%%%%%%%%%%%%%%%%%%%%%%

It is clearly visible from Figure~\ref{fig7} that the normalised gain varies proportionally with the divider current. In contrast to normalised gain, the normalised energy resolution exhibits the opposite trend. 

%%%%%%%%%%%%%%%%%%%%%%%%%%%%%%%%%%%%%%%%%%%%%%%%%%%
%\begin{figure}[htb!]
%	\begin{center} 
%	\includegraphics[scale=0.4]{norm_gain_current_fit.pdf}
%        \caption{Normalised gain as a function of bias current.}\label{fig11}
%	\end{center}
%\end{figure}
%%%%%%%%%%%%%%%%%%%%%%%%%%%%%%%%%%%%%%%%%%%%%%%%%%%

%%%%%%%%%%%%%%%%%%%%%%%%%%%%%%%%%%%%%%%%%%%%%%%%%%%
\begin{figure}[htb!]
	\begin{center} 
	\includegraphics[scale=0.5]{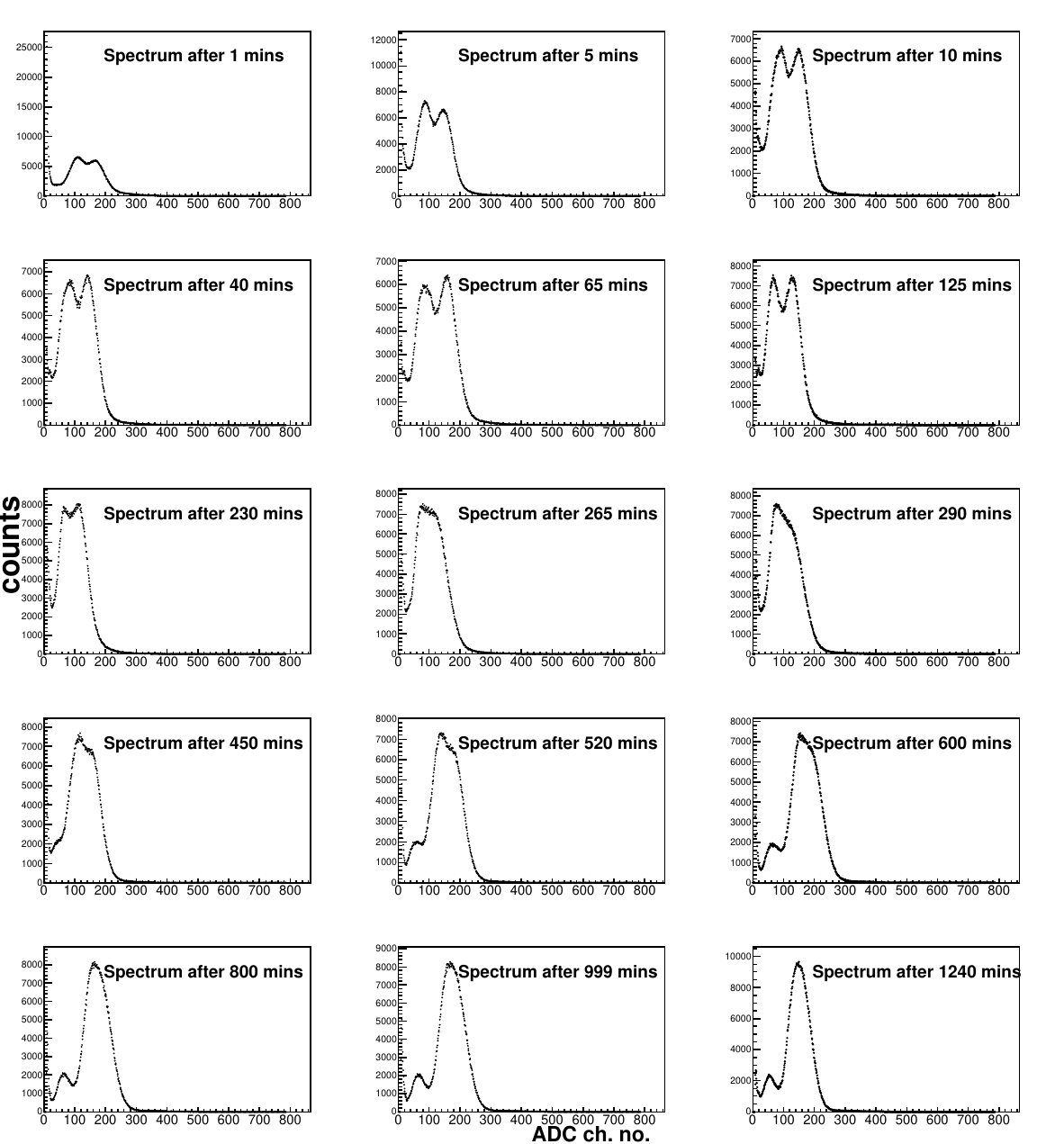}
	\caption{Variation of the shape of $^{55}$Fe X-ray spectra after different time from the application of HV. Appearance of an abnormal double peak at the beginning.}\label{fig8}
	\end{center}
\end{figure}
%%%%%%%%%%%%%%%%%%%%%%%%%%%%%%%%%%%%%%%%%%%%%%%%%%%

Another observation, which is much more significant for this study is the transition of deformed spectra to good spectra. It is observed from Figure~\ref{fig9} that, when the detector is first biassed to the desired set value the gain is too low and the spectrum is bad. After 3 minutes the gain increases and the Argon escape peak also started to appear. However, after 8 minutes a peculiar pattern in the $^{55}$Fe spectra is observed. It is noticed that the main peak of the spectrum is quite wide for about 26 minutes. Subsequently the spectra improved and 5.9~keV main peak along with the escape peak is observed after 55 minutes. It is also observed that with time as the current decreases the voltage across GEM foils ($\Delta$V) decreases and the gain also decreases and the peak moved towards left. So the nature of spectra from 8 to 26 minutes is not understood properly. 

Another example is given here for initial conditioning phase of the detector. In this case the HV is set at -~4500~V. After applying the voltage initially the spectra is saved in every 1 minute. Initially spectra are obtained with double peak structure. Well defined normal $^{55}$Fe spectra having one main peak and one escape peak are not obtained for about 10 hours. The change in shape of spectra are shown in Figure~\ref{fig8}.

Recently another peculiar behaviour is observed for this particular chamber. When bias voltage is applied then immediately the well separated mean peak and escape peak are not appearing. In fact the desired HV, say -~4600~V, can not be applied to the chamber directly. First HV need to set at a lower value, say -~4350~V, and then after each 60 minutes the bias voltage is increased by 100~V. Finally, around -~4600~V the well known $^{55}$Fe spectrum with a 5.9~keV main peak and an Argon escape peak is observed. If the voltage is raised to -~4600~V directly it started sparking inside the chamber with sound. However, such a peculiar behaviour is only observed when the voltage is raised to the chamber after a long period of shutdown. But if the power is suddenly shut down during a long-term continuous test of the detector and bias voltage is applied to GEM again within 1 hour, the above events does not occur. We are still trying to understand this behaviour of the chamber.
 
%%\label{}
%\lipsum[2]

%\subsection{Subsection title}
%\lipsum[3]

%\begin{table}
%\begin{tabular}{l c c c} 
% \hline
% Source & RA (J2000) & DEC (J2000) & $V_{\rm sys}$ \\ 
%        & [h,m,s]    & [o,','']    & \kms          \\
% \hline
% NGC\,253 & 	00:47:33.120 & -25:17:17.59 & $235 \pm 1$ \\ 
% M\,82 & 09:55:52.725, & +69:40:45.78 & $269 \pm 2$ 	 \\ 
% \hline
%\end{tabular}
%\caption{Random table with galaxies coordinates and velocities, Number the tables consecutively in
%accordance with their appearance in the text and place any table notes below the table body. Please avoid using vertical rules and shading in table cells.
%}
%\label{Table1}
%\end{table}

\section{Summary and discussion}
A comprehensive investigation is conducted on the study of stability of the gain and energy resolution of a SM triple GEM detector under continuous high-rate $^{55}$Fe X-ray irradiation using NIM electronics and a mixture of Ar and CO$_2$ in 70/30 volume ratio. In this continuous study it is observed that the T/p normalised gain decreases with time after initial charging up phase. The probable reason is that, although there is a sudden jump in the bias current observed at the starting, after sometimes the bias current started decreasing gradually. As a result the $\Delta$V across the GEM foils decrease which in turn reduce the gain of the detector. 

In addition to that several behavioural changes are also observed in the chamber. After a continuous operation of the chamber if there is a gap in operation and when the HV is ramped up then shape of some initial spectra are not normal. It has been observed that the time taken for the spectra to come to its normal shape can take up to 10-12 hours. Actually this particular chamber is under several long-term test since 2015. Several interesting results are being achieved from this particular chamber. During the charging-up studies good spectrum having main and escape peaks are obtained immediately after application of high voltage. However, since last year the bias current is decreasing with time which might be due to the ageing in the resistors of divider chain. The take home message from this article is that, although it is found there is no ageing in the chamber, and the decrease in normalised gain is directly related to the decrease in bias current, but still something happens to the chamber for using prolonged radiation for a few years so that immediately after applying HV the detector is not working properly, it is taking some time for conditioning, even the desired HV can not be applied directly without spark. The detail investigation is ongoing.
%%\label{}
%%\lipsum[4]

%\section{Summary and conclusions}
%%\label{}
%\lipsum[1-4]

%\section*{Acknowledgements}
\section{Acknowledgements}
	
	The authors would like to thank the RD51 collaboration for the support in building and initial	testing of the chamber in the RD51 laboratory at CERN. We would also like to thank Mr. Subrata Das for helping in building of the collimators used in this study. This work is partially supported by the research grant SR/MF/PS-01/2014-BI from DST, Govt. of India, and the research grant of the CBM-MuCh project from BI-IFCC, DST, Govt. of India. S. Mandal acknowledges his UGC-NET fellowship for the support. A. Sen acknowledges the support of the DST-INSPIRE Fellowship (DST/INSPIRE Fellowship/2018/IF180361). S. Biswas acknowledges the support of the DST-SERB Ramanujan Fellowship (D.O.No. SR/S2/RJN-02/2012).

%\appendix

%\section{Appendix title 1}

%\section{Appendix title 2}

%\bibliographystyle{elsarticle-harv} 
%\bibliography{example}
%\bibliography{}


\end{document}